\documentclass{article}

\usepackage{arxiv}

\usepackage[utf8]{inputenc} % allow utf-8 input
\usepackage[T1]{fontenc}    % use 8-bit T1 fonts
\usepackage{hyperref}       % hyperlinks
\usepackage{url}            % simple URL typesetting
\usepackage{booktabs}       % professional-quality tables
\usepackage{amsfonts}       % blackboard math symbols
\usepackage{nicefrac}       % compact symbols for 1/2, etc.
\usepackage{microtype}      % microtypography
\usepackage{lipsum}
\usepackage{color}
\usepackage{algorithm}
\usepackage{algorithmic}
\usepackage{amsmath,amsfonts,amssymb}
\usepackage{subfigure}
\usepackage{lineno}
\usepackage{graphicx}
\usepackage{bm}
\graphicspath{{figures/}}
\usepackage{subfigure}

\title{The effect of Reynolds number on the separated flow over a low-aspect-ratio wing}

\author{Luke Smith\thanks{Corresponding author: lsmith1@ucla.edu}   ,
\hspace{0.05em}
Kunihiko Taira
\\
Department of Mechanical and Aerospace Engineering, \\
University of California, Los Angeles, CA 90095, USA
}

\begin{document}
\maketitle

\begin{abstract}
At high incidence, low-aspect-ratio wings present a unique set of aerodynamic characteristics, including flow separation, vortex shedding, and unsteady force production. Furthermore, low-aspect ratio wings exhibit a highly impactful tip vortex, which introduces strong spanwise gradients into an already complex flow. In this work, we explore the interaction between leading edge flow separation and a strong, persistent tip vortex over a Reynolds number range of $600 \leq Re \leq 10,000$. In performing this study, we aim to bridge the insight gained from existing low Reynolds number studies of separated flow on finite wings ($Re \approx 10^2$) and turbulent flows at higher Reynolds numbers ($Re \approx 10^4$). Our study suggests two primary effects of Reynolds number. First, we observe a break from periodicity, along with a dramatic increase in the intensity and concentration of small-scale eddies, as we shift from $Re = 600$ to $Re = 2,500$. Second, we observe that many of our flow diagnostics, including the time-averaged aerodynamic force, exhibit reduced sensitivity to Reynolds number beyond $Re = 2,500$, an observation attributed to the stabilizing impact of the wing tip vortex. This latter point illustrates the manner by which the tip vortex drives flow over low-aspect-ratio wings, and provides insight into how our existing understanding of this flowfield may be adjusted for higher Reynolds number applications.
\end{abstract}

% ----------------------------------------------------------------------------------------------------------------------------------------------------------------------------

\section{Introduction}
\label{sec:introduction}

\noindent Low-aspect-ratio wings pose a unique aerodynamic challenge. Often employed in small-scale flight applications, including package delivery~\cite{golubev2012} and disaster relief~\cite{mohddaud2022}, low-aspect-ratio wings are expected to operate in conditions that break the classical assumption of inviscid, attached flow. These wings are thus prone to a host of unsteady, often undesirable aerodynamic effects, including flow separation, vortex shedding, and nonlinear oscillations in aerodynamic force~\cite{shyy2007}. At the same time, the compact footprint of these vehicles augments the impact of the wing tip vortex, imparting a large degree of downwash over a significant portion of the wing span~\cite{tang2004}. The resulting flowfield is tangled and complex: flow separation leads to the shedding of spanwise vortices, and these shed vortices undergo an intricate interaction with the wing tip. Such a three-dimensional, vortex-dominated flowfield precludes accurate prediction with conventional aerodynamic theory~\cite{bird2021}.

Partially because of this gap in understanding, there exists a growing body of literature concerned with the separated flow over aerodynamic bodies, with particular attention paid to the role of three-dimensionality. Many of these studies focus on spanwise periodic (or nominally two-dimensional) wings, and have produced detailed descriptions of the airfoil wake as a function of wing incidence, Reynolds number, and planform geometry~\cite{menon2020,kurtulus2021,jean2022}. Collectively, these studies suggest that a separated wake exhibits a limit cycle oscillation up to a critical Reynolds number of $Re \approx 1000$, at which point spanwise instabilities initiate the route to chaos~\cite{hoarau2003, zhang2009, he2017}. While rich with fundamental insight, studies of spanwise periodic wings necessarily lack the presence of a wing tip, and thus have limited applicability to fully three-dimensional, low-aspect-ratio-wings.

The simulation of finite wings, or those with a defined wing tip geometry, are more directly relevant to modern, small-scale flight application. In turn, computational efforts have begun to parse the interaction between a massively separated flow and a strong tip vortex. Taira and Colonius \cite{taira2009} were among the first to explore this phenomena in the low Reynolds number regime, performing direct numerical simulation (DNS) over a sweep of aspect ratios and incidence angles at $Re = 300$. The authors found that the tip vortex, by virtue of its downwash, imparts a stabilizing effect on leading edge separation, while simultaneously promoting complex linkages among wake vortices. Subsequent studies would expand the parameter space significantly, partitioning the flow into steady, periodic, and aperiodic regimes~\cite{zhang2020, pandi2023}, while also laying the groundwork for control-oriented descriptions of the flow~\cite{anton2022,jean2023}. 

The current work aims to address a persistent limitation in the existing literature: each study mentioned above was performed at a relatively low Reynolds number, with many studies limited to $Re < 1,000$. In this work, we aim to understand how a change in Reynolds number affects the physical interplay between tip vortex and leading edge separation. We also aim to bridge the gap between studies of finite wings at low Reynolds number, for which substantial progress has been made in the realm of stability and control, and studies of finite wings at higher, more turbulent Reynolds numbers~\cite{devenport1996, garmann2017}. We accomplish these goals by performing high fidelity computation over a sweep of Reynolds numbers in the range of $600 \leq Re \leq 10,000$. The following sections will address how these simulations were performed, and how the resulting flowfields differ from their low Reynolds number counterparts.

% ---------------------------------------------------------------------------------------------------------------------------------------------------------------------------------

\section{Methodology}
\label{sec:methodology}

\begin{figure}
\centerline{\includegraphics[width=0.95\textwidth]{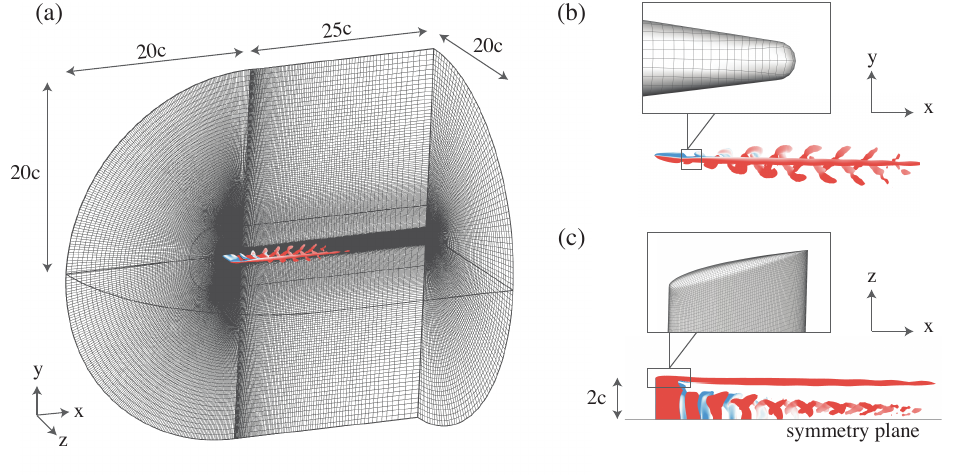}}%
\caption{(a) the computational domain, (b) trailing edge curvature, and (c) wing tip geometry for the present simulations.}
\label{fig1}
\end{figure}

In this work, we consider high-fidelity numerical simulations of a finite, NACA 0012 wing at five values of the chord-based Reynolds number, $Re = U_{\infty}c/\nu$. Figure~\ref{fig1} provides an isometric view of our computational domain and serves as a problem statement for the current work. In this figure, we define the streamwise ($x$), vertical ($y$), and spanwise ($z$) directions using Cartesian coordinates, and we place the origin $(0,0,0)$ at the leading edge of the wing root. Figure~\ref{fig1} also highlights several features of the wing surface geometry. For all simulations, we consider a semi aspect ratio $sAR = 2$ wing, outfitted with a rounded wing tip and a rounded trailing edge, at an incidence of $\alpha = 14^{\circ}$. We choose this specific combination of incidence and aspect ratio as a way of ensuring that both leading edge separation and vertical downwash are present over a substantial portion of the wing planform.

For all five Reynolds numbers, we perform computations using the compressible flow solver \textit{CharLES}, a finite volume solver with second-order accuracy in space and third-order accuracy in time~\cite{bres2017}. We formulate each simulation as either a direct numerical simulation ($Re \leq 2500$), or a large-eddy simulation ($Re \geq 5000$), with the Vreman subgrid scale model providing turbulent closure for our large-eddy simulations~\cite{vreman2004}. In both formulations, we prescribe a Dirchlet boundary condition ($U_\infty/a_{\infty} = 0.1$) at the inlet and farfield boundaries; a sponge boundary condition (with a spatial window of $x/c \in [15,25]$) at the outlet boundary; and an adiabatic wall boundary condition at the airfoil surface. We impose a symmetry condition at the wing mid-span ($z = 0$) as a means of minimizing our cell count. When advancing the simulation in time, we select a time-step such that the local Courant number remains below $U_{\infty} \Delta t / \Delta x = 1$ throughout the entire domain.

Because we are interested in the separated flow regime, our flowfield will exhibit characteristics similar to bluff body vortex shedding, and we expect to observe an increasingly broad spectrum of flow structures as we increase the Reynolds number. We thus generated multiple volumetric grids for this study to ensure sufficient resolution of the energy-containing scales. We build each mesh by first generating a structured grid along the airfoil surface, extending this grid as a two-cell-thick block along the streamwise extent of the wake, and space-marching the combined grid outward along the surface normal. We choose the resolution of our surface mesh, along with the growth rate of the space-marching procedure, such that each simulation is considered either a DNS ($Re \leq 2,500$) or an LES ($Re \geq 5,000$). Additional details regarding our choice of grid resolution, along with a quantification of the grid-dependence inherent to our results, can be found in appendix~\ref{app}.

% ---------------------------------------------------------------------------------------------------------------------------------------------------------------------------------

\section{Results}
\label{sec:results}

\begin{figure}
\centerline{\includegraphics[width=0.975\textwidth]{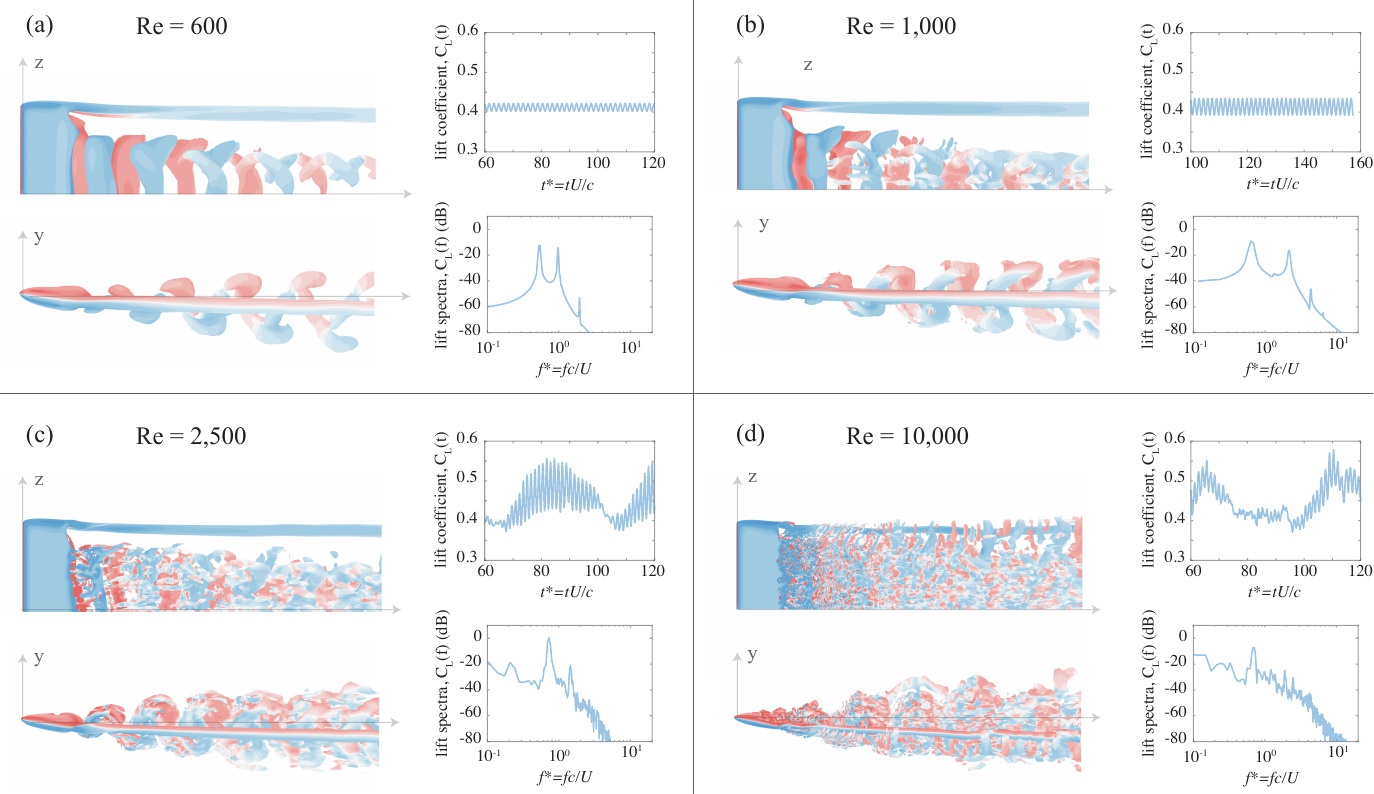}}%
\caption{Instantaneous isocontours of Q-criterion, lift coefficient trace, and lift spectra for (a) $Re = 600$, (b) $1,000$, (c) $2,500$ and (d) $10,000$.}
\label{fig2}
\end{figure}

We now examine the flowfield produced by a low-aspect-ratio wing ($sAR = 2$) at constant incidence ($\alpha = 14^{\circ}$) over a sweep of Reynolds numbers ($600 \leq Re \leq 10,000$). We begin by considering a qualitative overview of the flow as a function of Reynolds number, before focusing on the interaction between the tip vortex and shed wake. Figure~\ref{fig2} shows a series of flowfield metrics intended to capture the scale, character, and intensity of unsteady flow structures across our sweep of Reynolds numbers. Each subpanel of figure~\ref{fig2} includes three representations of the unsteady flowfield: an instantaneous snapshot of Q-criterion, colored by contours of spanwise vorticity ($\omega_z$); a time history of the lift coefficient, integrated along the entire body of the wing; and a spectral description of the unsteady lift coefficient, computed via Welch's method. Note that when computing the lift spectra, we only include measurements collected at least 45 convective times beyond the simulation's impulsive start.

Looking at figure~\ref{fig2}, we observe several striking changes in the character of the flow as we increase Reynolds number. We begin by noting that the $Re = 600$ case appears organized and coherent: we can visually identify shed vortices, the lift coefficient evolves periodically in time, and the lift spectra is dominated by a single vortex shedding frequency (the Strouhal number) and its harmonics. The flow behaves similarly at $Re = 1,000$, with only a slight increase in the magnitude of lift oscillations. As we move from $Re = 1,000$ to $Re = 2,500$, the flow undergoes a notable change, losing much of its qualitative coherence. While we can still identify shed vortices in the near-wing region, the wake at $Re = 2,500$ is interspersed with small-scale shear linkages, and the lift spectra transitions to a more broadband distribution (while still exhibiting a minor peak at the Strouhal shedding frequency). This breakdown of the wake is intensified at $Re = 10,000$, at which point shed vortices are almost entirely obscured by small-scale eddies.

\begin{figure}
\centerline{\includegraphics[width=0.975\textwidth]{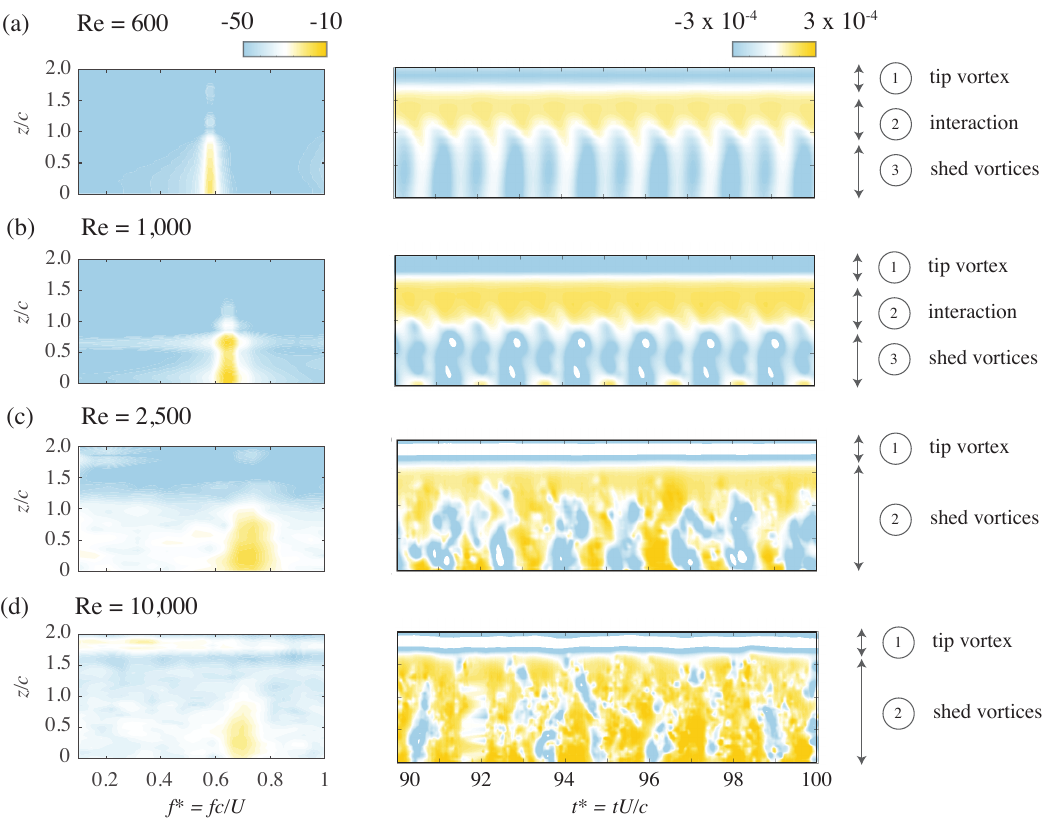}}%
\caption{(left) the pressure spectra and (right) instantaneous pressure at $(x/c, y/c)$ = $(3.0, -0.25)$ for (a) $Re = 600$, (b) $1,000$, (c) $2,500$ and (d) $10,000$.}
\label{fig3}
\end{figure}

Figure~\ref{fig3} offers an alternate view of the the airfoil at multiple Reynolds numbers. In this figure, we plot the signal from a pressure rake positioned downstream of the airfoil surface; the rake consists of 50 spanwise-aligned pressure probes at $(x/c,y/c) = (3.0,-0.24)$. The right column of figure~\ref{fig3} shows the temporal evolution of this pressure signal, while the left column describes its spectral composition as a function of spanwise position ($z/c$). 

Echoing the flowfield snapshots of figure~\ref{fig2}, we observe a growing incoherence in the pressure signal as Reynolds number increases, an incoherence that appears to increase sharply between $Re = 1,000$ and $2,500$. Whereas figures~\ref{fig3}(a)~and~(b) exhibit clear spikes that can be linked to distinct flow structures (i.e., shed vortices, a tip vortex, and a quasi-steady region in which the two interact), the remaining rows of figure~\ref{fig3} suggest a breakdown of these flow structures. Figure~\ref{fig3}(c) shows the emergence of ``choppiness'' in the structure of each pressure band, pointing toward the irregular, three-dimensional nature of passing vortices, whereas figure~\ref{fig3}(d) sees this irregularity spread outboard, into the region of tip vortex interaction. The spectral character of the wake is also impacted by the increase in Reynolds number. The left column of figure~\ref{fig3} reveals that the wake's frequency content shifts to a higher frequency and dims in intensity, implying that the signal has drifted into aperiodicity.    

Altogether, figures~\ref{fig2}~and~\ref{fig3} suggest the presence of two vortex shedding regimes. In the first regime ($600 \leq Re \leq 1,000$), the airfoil wake exhibits strong periodicity and its spatial structure can be partitioned into distinct regions. In the second regime ($2,500 \leq Re \leq 10,000$), these spatial regions break down, the structure of the wake becomes disorganized, and broadband unsteadiness spreads from wing root to wing tip. 

We now focus our attention on a specific aspect of the flow over a finite wing: the interplay between shed vortices, or those resulting from leading edge separation, and the tip vortex that forms at the rounded edge of our wing. This interaction is key to the onset of wake disorganization; while itself a quasi-steady flow feature, the tip vortex plays a crucial role in the onset of shear instability and vortex rollup~\cite{huang1995,jean2022}. We begin by visualizing the effect of Reynolds number on the leading-edge shear layer before reframing these effects in the context of the tip vortex.

 \begin{figure}
\centerline{\includegraphics[width=\textwidth]{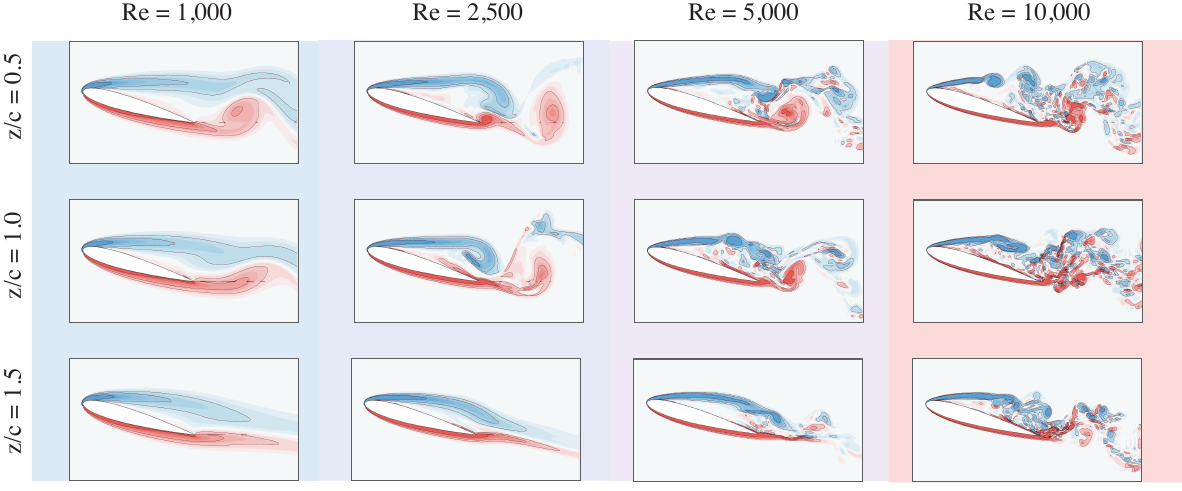}}%
\caption{Slices of the spanwise vorticity field ($\omega_z$) for a sweep of Reynolds numbers (rows) and spanwise locations (columns).}
\label{fig4}
\end{figure}

Figure~\ref{fig4} shows instantaneous slices of the the spanwise vorticity field ($\omega_z$) collected in the near-wing portion of our computational domain. We organize this figure such that $Re$ increases along each row, while $z/c$ moves outboard along each column. Let us begin by considering the first row of figure~\ref{fig4}, which shows the unsteady flow at $z/c = 0.50$. Looking specifically at the flow near the leading edge, we observe a clear trend: as we increase the Reynolds number, the leading edge shear layer becomes thinner, and the onset of shear layer instability moves upstream. The $Re = 1,000$ case exhibits only a minor accumulation of negative vorticity as it interacts with the trailing edge vortex; by $Re = 2,500$, the flow exhibits a clear rollup in a region just downstream of the trailing edge; by $Re = 10,000$, small-scale vortex rollup is visible as early as the airfoil mid-chord. The first row of figure~\ref{fig4} suggests that an increase in Reynolds number promotes the upstream rollup of spanwise vortices, which leaves the near-wing region prone to disorganization.

Next, let us examine how these vortex structures change as we move closer to the wing tip. Consider the second column of figure~\ref{fig4}, which shows the vorticity field ($\omega_z$) at $Re = 2,500$ over a sweep of spanwise locations. In the first row of this column, we observe the expected instability of the leading edge shear layer, with vortex rollup occurring near the trailing edge. As we move down the column, we observe attenuation of this instability. In fact, the flowfield at $z/c = 1.5$ exhibits a relatively stable shear layer for $Re = 2,500$, with minimal indication of vortex rollup or complex unsteadiness. Moving to $Re = 10,000$, vortex rollup is visible just beyond the midchord for all spanwise stations, but the degree of disorganization is substantially more mild at $z/c = 1.5$ compared to regions closer to the root. In this sense, figure~\ref{fig4} suggests that the tip vortex imparts a stabilizing effect on the leading edge shear layer, an effect that grows more significant as we approach the wing tip. Such an observation can be attributed to the role of downwash in an aerodynamic flow: as we move closer to the tip, downwash becomes more severe and local incidence becomes more mild, ultimately leading to an attenuation of instability in the separated shear layer.

\begin{figure}
\centerline{\includegraphics[width=\textwidth]{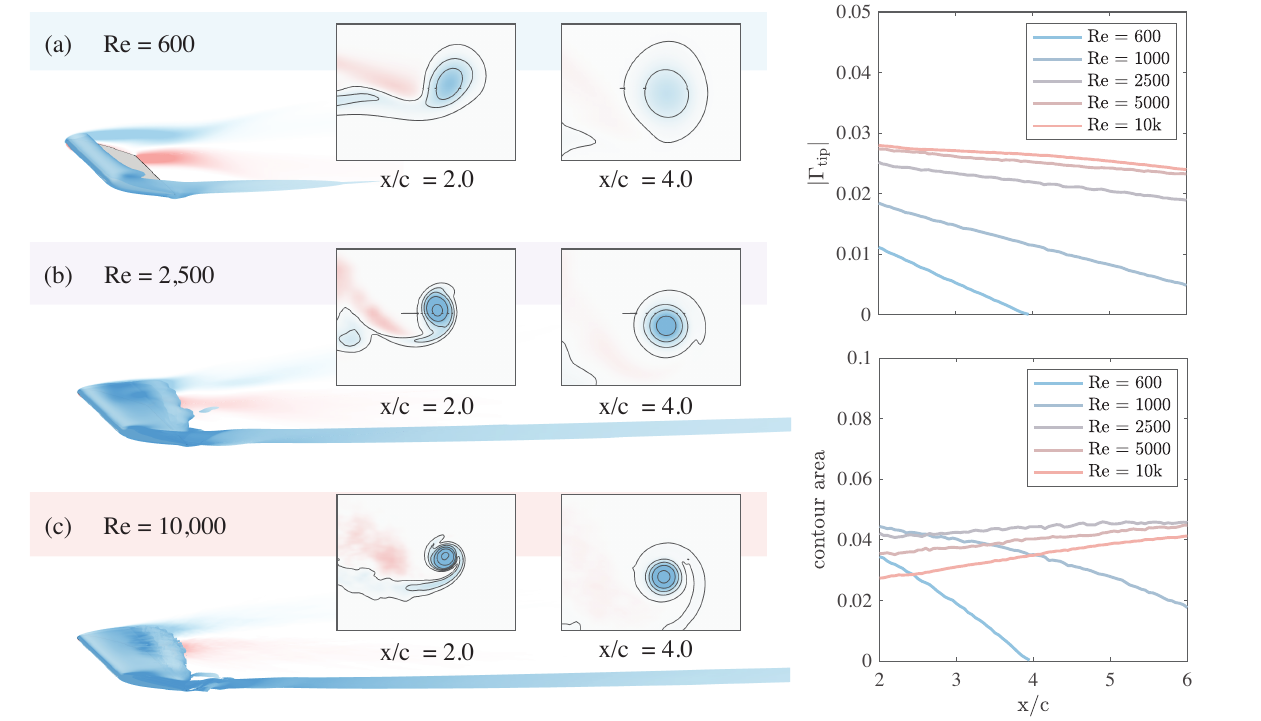}}%
\caption{Spatial evolution of the tip vortex in the time-averaged flowfield for (a) $Re = 600$, (b) $2,500$, and (c) $10,000$.}
\label{fig5}
\end{figure}

Broadly, figure~\ref{fig4} demonstrates that the local structure of the shed wake is strongly dependent on its proximity to the tip vortex, an observation that holds across all Reynolds numbers in the current study. Figure~\ref{fig5} expands upon this idea by quantifying the strength of the tip vortex. In this figure, we plot time-averaged isocontours of Q-criterion, along with planar snapshots of the streamwise vorticity field ($\omega_x$), for a sweep of Reynolds numbers over $600 \leq Re \leq 10,000$. We also track the circulation and planar area associated with a single isocontour; these two quantities are plotted on the right-hand side of figure~\ref{fig5}. For both methods of visualization, we observe a similar trend: as we increase the Reynolds number, the tip vortex grows in intensity and persistence, exhibiting reduced spatial dissipation beyond $Re = 1,000$. This observation is supported by the left-hand side of figure~\ref{fig5}, which exhibits more concentrated, higher intensity contours as the Reynolds number increases, and by the right-hand side of figure~\ref{fig5}, which shows that the tip vortex maintains its strength and size for a substantially longer streamwise extent at Reynolds numbers beyond $Re = 1,000$.

We thus arrive at one of the main conclusions of this work. That is, an increase in Reynolds number produces two seemingly opposing effects: (1) an increase in the likelihood of shear layer rollup for inboard regions of the wing, and (2) an increase in the strength and persistence of the tip vortex, which attenuates vortex rollup in outboard regions of the wing. This dichotomy provides a simple explanation for the regime changes observed in figure~\ref{fig2}. At low Reynolds number ($Re \leq 1,000$), the degree of shear instability is mild, and the strength of the tip vortex is enough to suppress vortex breakdown across the entirety of the wing span. Between $Re = 1,000$ and $2,500$, the degree of instability grows to the point that inboard portions of the wing ($0 \leq z/c \leq 1.5$) transition to irregular, unsteady shedding, while outboard portions of the wing ($0.50 \leq z/c \leq 0.0$) remain quasi-steady. By $Re = 10,000$, the degree of shear instability outpaces the growing strength of the tip vortex, and small-scale vortex structures begin to emerge near the tip vortex, albeit gradually. Note that while outboard unsteadiness has just begun to emerge by $Re = 10,000$, we expect the general structure of figure~\ref{fig2} to persist beyond $Re = 10,000$. Previous studies~\cite{garmann2017, toosi2024} have reported a quasi-steady tip vortex, coupled with the widespread entrainment of small-scale vortex structures, at Reynolds numbers $O(10^5)$, suggesting that the gradual outboard spread of unsteadiness continues to higher $Re$.

\begin{figure}
\centerline{\includegraphics[width=\textwidth]{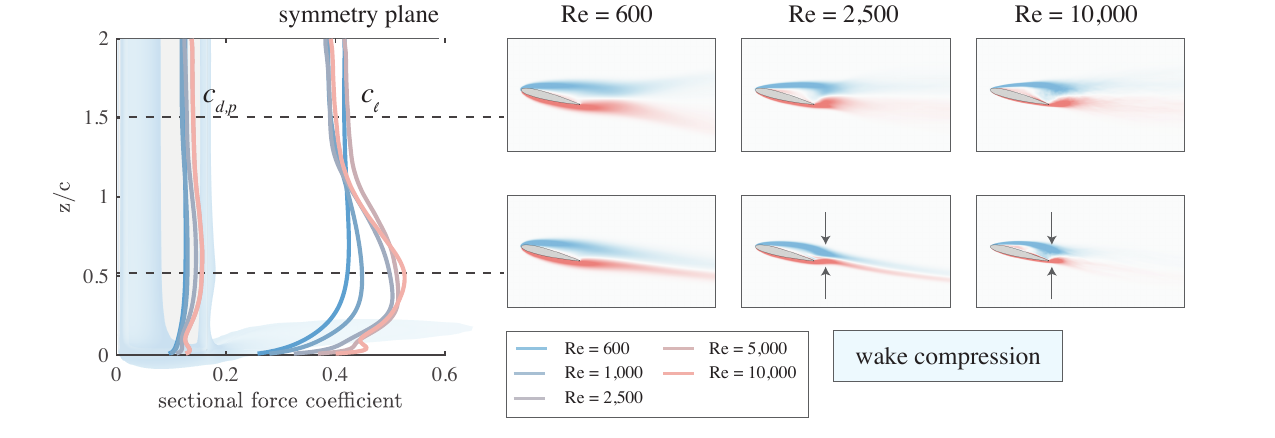}}%
\caption{Spanwise distribution of the time-averaged lift coefficient ($c_l$) and pressure drag coefficient ($c_{d,p}$) for a sweep of Reynolds numbers. Insets show the mean spanwise vorticity field at specific spanwise locations.}
\label{fig6}
\end{figure}

The trend described above, wherein the tip vortex stabilizes outboard portions of the wing, is particularly relevant to wings with a short span, and has important implications on the production of aerodynamic force. Figure~\ref{fig6} plots the time-averaged lift and pressure-drag coefficients as a function of span for a sweep of Reynolds numbers, with insets showing the time-averaged vorticity field ($\omega_z$) at select locations. In this figure, we observe a large, outboard spike in sectional lift coefficient for Reynolds numbers beyond $Re = 600$. While tip vortices are conventionally associated with a reduction in sectional lift, the presence of this spike is consistent with many of our observations thus far. For example, if we move along the bottom row of figure~\ref{fig6}, we observe that vorticity is pulled closer to the airfoil surface with increasing Reynolds number, thus strengthening its contribution to vortex lift~\cite{lee2012}. We can in turn view the spanwise lift distribution as a reflection of the interplay between vortex rollup and local downwash; that is, shear instabilities concentrate vorticity into discrete structures, while downwash draws these structures closer to the wing surface. 
 
Up to this point, we have limited our discussion of low-aspect-ratio wings to the interaction between the tip vortex and the leading edge shear layer. While critical to the production of aerodynamic force, this interaction is constrained to the near-wing portion of the domain, and is only implicitly related to the behavior of vortices as they move downstream. Thus, we next consider the three-dimensional structure of shed vortices over many convective times, augmenting our portrait of the airfoil wake with the details of its downstream structure.

Because our parameter space includes higher Reynolds numbers ($Re > 1,000$), it can be difficult to visualize the structure of wake vortices via traditional techniques. We must instead rely upon some means of temporal, conditional averaging in order to parse the complex interactions that characterize the downstream wake. Here, we employ the Dynamic Mode Decomposition (DMD) to accomplish this task. DMD operates by finding the best-fit linear operator for the discrete dynamical system governing the evolution of the flow; the eigenvectors of this linear operator are called ``modes'' and represent recurrent spatial patterns in a time-series of flowfield snapshots~\cite{kutzDMDBook}. Because DMD assumes a linear system, each of these modes can be linked to a discrete frequency. 

\begin{figure}
\centerline{\includegraphics[width=\textwidth]{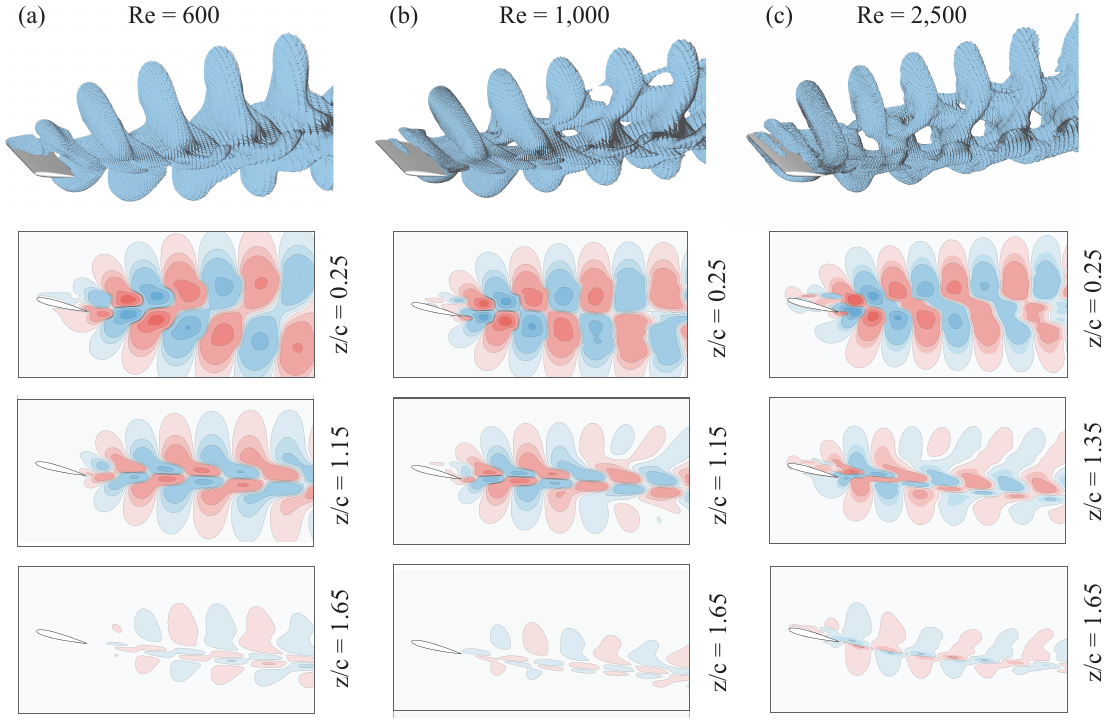}}%
\caption{The three-dimensional DMD mode associated with the dominant vortex shedding frequency for (a) $Re = 600$, (b) $Re = 1,000$, and (c) $Re = 2,500$.}
\vspace{-0.15em}
\label{fig7}
\end{figure}

With this in mind, figure~\ref{fig7} shows the dominant DMD mode for three representative Reynolds numbers, computed using streamwise velocity ($u_x$) as the state variable. For each case in this figure, we select the appropriate eigenvector by extracting a dominant vortex shedding frequency from the unsteady lift spectra (see figure~\ref{fig2}) and matching this frequency with the nearest (purely imaginary) eigenvalue from DMD. Note that because the flow's aperiodicity grows with Reynolds number, we require an increasing number of snapshots to ensure convergence at higher Reynolds numbers. We thus incorporate 200 snapshots in the computation of the $Re = 600$ mode; 300 snapshots in the computation of the $Re = 1,000$ mode; and 2,000 snapshots for the computation of the $Re = 2,500$ mode.

Let us begin by considering the DMD mode associated with our lowest Reynolds number. In figure~\ref{fig7}(a), we observe a sequence of tube-like structures cascading through the wake of the airfoil, a common modal pattern associated with periodic vortex shedding and convection. Each of these tube-like structures resembles an \textit{arch} in three-dimensional space. If we shift our attention downstream, these arches undergo substantial tilting and distortion over time, with outboard regions transitioning from a spanwise ($z$) alignment to more of a streamwise ($x$) alignment. The timing, behavior, and distortion of these tubes is closely linked to impact of the tip vortex on the shed wake. Specifically, the arch-like structure of each tube is indicative of a strong spanwise gradient in downwash velocity, as outboard regions of each tube are subject to intense downwash from the tip vortex. Meanwhile, the streamwise tilting of each tube is indicative of a strong spanwise gradient in streamwise velocity, as outboard regions of each tube are disrupted by the axial velocity of the tip vortex. 

We now examine how this arch structure changes with an increase in Reynolds number. Looking at figures~\ref{fig7}(b)~and~\ref{fig7}(c), we still observe the familiar chain of arch-like modal structures, but with two primary differences compared to $Re = 600$. First, we note that the degree of streamwise vortex tilting (i.e., the rate at which each ``tube'' is re-oriented toward the $x$-axis) appears to increase as we increase the Reynolds number. This observation can be linked to the behavior of the tip vortex described in figure~\ref{fig5}. Because the strength and persistence of the tip vortex increases directly with the Reynolds number, each tube is subject to a sharper, more intense spanwise gradient, which persists over an increasingly large streamwise portion of the wake. In turn, the modal structures of figures~\ref{fig7}(b)~and~\ref{fig7}(c) show signs of streamwise tilting at significantly more upstream locations compared to figure~\ref{fig7}(a). 

Second, we note that the ``legs'' of each tube, or the outboard regions in which each tube tilts rapidly toward the tip vortex, appear increasingly distinct as we move from figure~\ref{fig7}(a) to figure~\ref{fig7}(b). This observation implies that the ``interaction'' region, over which shed vortices merge with the tip vortex, becomes increasingly compressed as we increase the Reynolds number. Such an observation aligns with the physics of the airfoil wake; in figure~\ref{fig3}, we observed that the bounds of the ``interaction'' region are pushed outboard as wake disorganization casts a wider breadth.

All in all, the DMD modes of figure~\ref{fig7} indicate that an increase in Reynolds number produces two primary changes to the downstream structure of shed vortices: (1) an increase in the intensity of streamwise vortex tilting, and (2) an outboard shift in the region where this tilting is concentrated. These trends are consistent with our physical observations regarding the strength and persistence of the tip vortex, and demonstrate the close coupling between shed vortices and tip vortices observed across the current Reynolds number range.

% ---------------------------------------------------------------------------------------------------------------------------------------------------------------------------------

\section{Concluding Remarks}

In this work, we investigated the unsteady flow over a low-aspect-ratio ($sAR=2$) wing at multiple values of the freestream Reynolds number ($600 \leq Re \leq 10,000$). Our broad goal was to characterize the airfoil wake as it transitions from large-scale, periodic vortex shedding ($Re \approx 10^2$) to disorganization and three-dimensionality ($Re \approx 10^4$), with an emphasis on the role of the wing tip vortex. Using a combination of direct numerical simulation ($Re \leq 2,500$) and large-eddy simulation ($Re \geq 5,000$), we identified a regime change near $Re = 1,000$, in which wake vortices break down, aerodynamic force becomes aperiodic, and disorganization overtakes inboard portions of the airfoil wake. For $Re > 1,000$, we find that an increase in Reynolds number promotes instability in the leading edge shear layer, while also strengthening the wing tip vortex, which implicitly suppresses instability in outboard regions of the wing. The result is a gradual spread of fine-scale structures from the inboard, nominally 2D portion of the wing, toward the outboard, tip-vortex-influenced portions of the wing. Collectively, these observations illuminate how a strong, persistent tip vortex impacts the structure of a transitioning shear wake, and lay the groundwork for a physical understanding of low-aspect-ratio wings in realistic, high Reynolds number conditions.

% ---------------------------------------------------------------------------------------------------------------------------------------------------------------------------------

\newpage
\section*{Acknowledgements}

LS and KT wish to thank Dr. Daniel Garmann for thoughtful discussion on the dynamics of separated, aerodynamic flows. LS and KT acknowledge the generous support of the US Air Force Office of Scientific Research (Grant No. FA9550-21-1-0174) and the US Department of Defense Vannevar Bush Faculty Fellowship (Grant No. N00014-22-1-2798). The authors report no conflict of interest.

% ---------------------------------------------------------------------------------------------------------------------------------------------------------------------------------

\appendix{}
\section{Grid resolution}\label{app}

\begin{table}
  \begin{center}
\def~{\hphantom{0}}
  \begin{tabular}{ccccccc}
      $Re$  & $n_{\text{airfoil}}$   &   $n_y$     &    $n_z$    &    $n_w$   &   $y_0^+$     &      cell count ($10^6$)  \\[5pt]
       600   & 80 & 160 & 50 & 170 & 0.17 & 4.0 \\
       1,000   & 120 & 200 & 75 & 260 &  0.14 & 11.4 \\
       2,500  & 120 & 200 & 75 & 260 & 0.33 & 11.4 \\
       5,000   & 160 & 250 & 105 & 345 & 0.31 & 26.5 \\
       10,000 & 160 & 250 & 105 & 345 & 0.59 & 26.5 \\ \\
  \end{tabular}
  \caption{Geometric resolution parameters for each of the five Reynolds number cases.}
  \label{table1}
  \end{center}
\end{table}

In this appendix, we provide additional details regarding the setup of our computational domain. We begin by noting that for all Reynolds numbers considered here, the resolution of our computational domain is governed by the following geometric parameters: the number of grid points allocated to the wing tangent ($n_{\text{airfoil}}$); the number of grid points allocated to the wing span ($n_z$); the number of grid points allocated to the wing normal ($n_y$); and the number of streamwise grid points allocated to the airfoil wake ($n_w$). Table~\ref{table1} assigns values to these parameters for each of our five Reynolds number cases; this table also reports the value of our initial off-wall spacing ($y_0^+$) in viscous wall units. We choose the value of each parameter such that the average edge length within the near-wing mesh (i.e., the region corresponding to $-2.0 \leq x/c, y/c, z/c \leq 2.0$) roughly conforms to a set of grid spacing limitations. For $Re \leq 2500$, these limitations correspond to a DNS, i.e, $(\Delta x^+,\Delta y^+,\Delta z^+) \leq (2.5, 2.5, 5.0)$, while for $Re \geq 5000$, these limitations correspond to an LES, i.e, $(\Delta x^+,\Delta y^+,\Delta z^+) \leq (5.0, 5.0, 10.0)$. Note that for both DNS and LES cases, we select an initial off-wall spacing such that the innermost region near the wall ($y^+ < 5$) contained a minimum of 6 grid points.

\begin{figure}
\centerline{\includegraphics[width=0.95\textwidth]{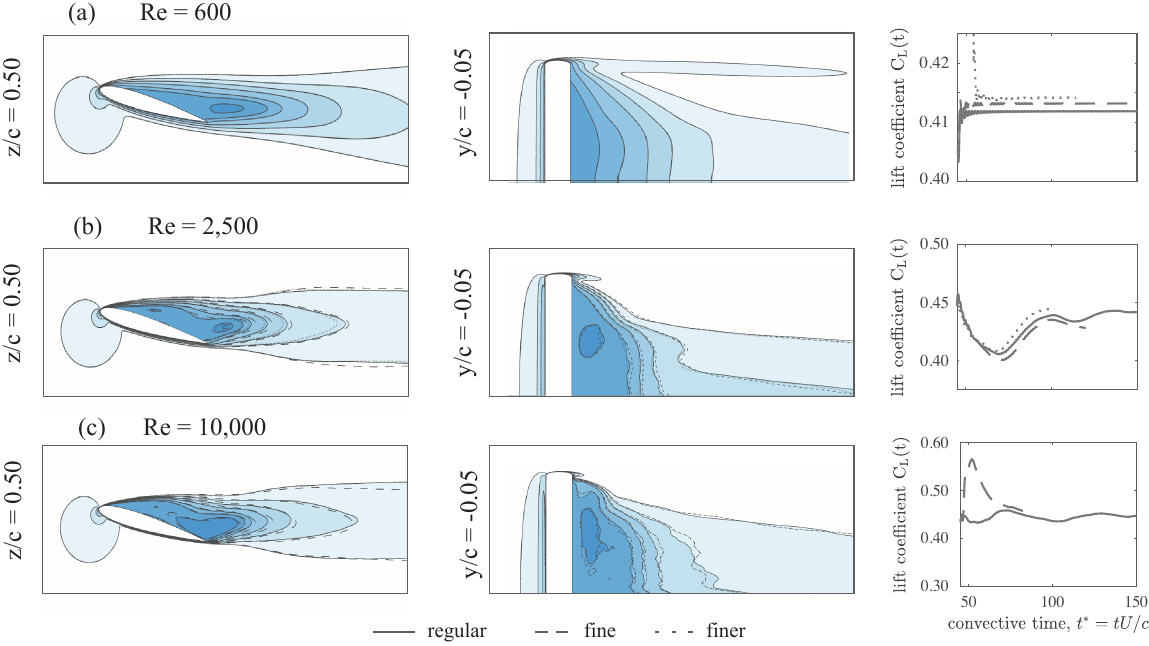}}%
\caption{The effect of grid resolution on the time-averaged, streamwise velocity field (left and center columns) and the time-averaged lift coefficient (right column).}
\label{fig8}
\end{figure}

Figure~\ref{fig8} collects the results of a grid-independence study undertaken to assess the accuracy of the mesh parameters reported above. In figure~\ref{fig8}, each subplot corresponds to a separate Reynolds number, and shows iso-contours of streamwise velocity, averaged over a temporal window of $>$ 50 convective time units. Each subplot also includes a running average of the lift coefficient, $C_L = 2 F_y/ \rho U_{\infty}^2 S$, as a means of visualizing the flow's global convergence behavior. For each Reynolds number, the label ``regular" corresponds to the level of grid refinement reported in table~\ref{table1}. When evaluating the grid independence of a given Reynolds number, we generate a new ``fine" and ``finer" grid by applying refinement uniformly across our four governing geometric parameters. Each subsequent level of grid refinement roughly doubles the total cell count of the mesh. Note that because of a prohibitively high cell count, we limit the $Re = 10,000$ case to a single iteration of grid refinement.

Figure~\ref{fig8} reveals that our time-averaged solution exhibits minimal sensitivity to successive levels of grid refinement. The iso-contour lines of figures~\ref{fig8}(a) and (b) are nearly identical to one another, suggesting that a ``regular" grid provides sufficient resolution for the $Re = 600$ and $Re = 2, 500$ cases. The downstream iso-contour lines begin to deviate from one another at $Re = 10, 000$, likely owing to the flow's increasing spatiotemporal irregularity, but the differences are slight, and even for $Re = 10,000$, the aerodynamic force appears to converge after a few dozen convective times. We thus consider our solution spatially converged; the results presented throughout this work correspond to the ``regular'' grids outlined in table~\ref{table1}. 

\bibliographystyle{unsrt}
\bibliography{manuscript}

\begin{thebibliography}{10}

\bibitem{golubev2012}
V.V. Golubev and M.R. Visbal.
\newblock Modeling {MAV} response in gusty urban environments.
\newblock {\em International Journal of Micro Air Vehicles}, 4(1):79--92, 2012.

\bibitem{mohddaud2022}
S.M.S.M. Daud, M.Y.P.M. Yusof, C.C. Heo, L.S. Khoo, M.K.C. Singh, M.S. Mahmood,
  and H.~Nawawi.
\newblock Applications of drone in disaster management: A scoping review.
\newblock {\em Science and Justice}, 62(1):30--42, 2022.

\bibitem{shyy2007}
Wei Shyy, Yongsheng Lian, Jian Tang, Dragos Viieru, and Hao Liu.
\newblock {\em Aerodynamics of low {R}eynolds number flyers}.
\newblock Cambridge University Press, 1 edition, 2007.

\bibitem{tang2004}
T.~Jian and Z.~Ke-Qin.
\newblock Numerical and experimental study of flow structure of
  low-aspect-ratio wing.
\newblock {\em Journal of Aircraft}, 41(5), 2004.

\bibitem{bird2021}
H.J.A. Bird and K.~Ramesh.
\newblock Unsteady lifting-line theory and the influence of wake vorticity on
  aerodynamic loads.
\newblock {\em Theoretical and Computational Fluid Dynamics}, 35:609--631,
  2021.

\bibitem{menon2020}
K.~Menon and R.~Mittal.
\newblock Aerodynamic characteristics of canonical airfoils at low {R}eynolds
  numbers.
\newblock {\em AIAA Journal}, 58(2):977--980, 2020.

\bibitem{kurtulus2021}
D.F. Kurtulus.
\newblock Vortex flow aerodynamics behind a symmetric airfoil at low angles of
  attack and {R}eynolds numbers.
\newblock {\em Physical Review E}, 79:045306, 2021.

\bibitem{jean2022}
J.H.M. Ribeiro, C.~Yeh, K.~Zhang, and K.~Taira.
\newblock Wing sweep effects on laminar separated flows.
\newblock {\em Journal of Fluid Mechanics}, 950:A23, 2022.

\bibitem{hoarau2003}
Y.~Hoarau, M.~Braza, Y.~Ventikos, D.~Faghani, and G.~Tzabiras.
\newblock Organized modes and three-dimensional transition to turbulence in the
  incompressible flow around a {NACA}0012 wing.
\newblock {\em Journal of Fluid Mechanics}, 496:63--72, 2003.

\bibitem{zhang2009}
N.~Liu J.~Zhang and X.~Lu.
\newblock Route to a chaotic state in fluid flow past an inclined plate.
\newblock {\em Physical Review E}, 79:045306, 2009.

\bibitem{he2017}
W.~He, R.S. Gioria, J.M. Perez, and V.~Theofilis.
\newblock Linear instability of low {R}eynolds number massively separated flow
  around three {NACA} airfoils.
\newblock {\em Journal of Fluid Mechanics}, 811:701--741, 2017.

\bibitem{taira2009}
K.~Taira and T.~Colonius.
\newblock Three-dimensional flows around low-aspect-ratio flat-plate wings at
  low {R}eynolds numbers.
\newblock {\em Journal of Fluid Mechanics}, 623:187--207, 2009.

\bibitem{zhang2020}
K.~Zhang, S.~Hayostek, M.~Amitay, W.~He, V.~Theofilis, and K.~Taira.
\newblock On the formation of three-dimensional separated flows over wings
  under tip effects.
\newblock {\em Journal of Fluid Mechanics}, 895:A9, 2020.

\bibitem{pandi2023}
J.S. Pandi and S.~Mittal.
\newblock Streamwise vortices, cellular shedding, and force coefficients on a
  finite wing at low {R}eynolds number.
\newblock {\em Journal of Fluid Mechanics}, 958:A10, 2023.

\bibitem{anton2022}
A.~Burtsev, W.~He, K.~Zhang, V.~Theofilis, K.~Taira, and M.~Amitay.
\newblock Linear modal instabilities around post-stall swept finite wings at
  low {R}eynolds number.
\newblock {\em Journal of Fluid Mechanics}, 944:A6, 2022.

\bibitem{jean2023}
J.H.M. Ribeiro, C.~Yeh, and K.~Taira.
\newblock Triglobal resolvent analysis of swept-wing wakes.
\newblock {\em Journal of Fluid Mechanics}, 954:A42, 2023.

\bibitem{devenport1996}
W.J. Devenport, M.C. Rife, S.I Liapis, and G.J. Follin.
\newblock The structure and development of a wing-tip vortex.
\newblock {\em Journal of Fluid Mechanics}, 312:67--106, 1996.

\bibitem{garmann2017}
D.J. Garmann and M.R. Visbal.
\newblock Analysis of tip vortex near-wake evolution for stationary and
  oscillating wings.
\newblock {\em AIAA Journal}, 55(8):2686--2702, 2017.

\bibitem{bres2017}
G.A. Br\`{e}s, F.E. Ham, J.W. Nichols, and S.K. Lele.
\newblock Unstructured large-eddy simulations of supersonic jets.
\newblock {\em AIAA Journal}, 55(4):1164--1184, 2017.

\bibitem{vreman2004}
A.~Vreman.
\newblock An eddy-viscosity subgrid-scale model for turbulent shear flow:
  algebraic theory and applications.
\newblock {\em Physics of Fluids}, 16:3670, 2004.

\bibitem{huang1995}
R.F. Huang and C.L. Lin.
\newblock Vortex shedding and shear-layer instability of wing at low-{R}eynolds
  numbers.
\newblock {\em AIAA Journal}, 33(8):1398--1403, 1995.

\bibitem{toosi2024}
S.~Toosi, A.~Peplinski, P.~Schlatter, and R.~Vinuesa.
\newblock The impact of finite span and wing-tip vortices on a turbulent
  {NACA}0012 wing, 2023.

\bibitem{lee2012}
J-J Lee, C-T Hsieh, C.C Chang, and C-C Chu.
\newblock Vorticity forces on an impulsively started flat plate.
\newblock {\em Journal of Fluid Mechanics}, 694:464--492, 2012.

\bibitem{kutzDMDBook}
J.N. Kutz, S.L. Brunton, B.W. Brunton, and J.L. Proctor.
\newblock {\em Dynamic mode decomposition: data-driven modeling of complex
  systems}.
\newblock Society for Industrial and Applied Mathematics, 1 edition, 2016.

\end{thebibliography}

\end{document}